\documentclass[prl,twocolumn,preprintnumbers,amsmath,amssymb,showpacs]{revtex4}
\usepackage{graphicx}
\usepackage{dcolumn}
\usepackage{bm}
\usepackage{mathrsfs}
\usepackage{subfigure}
\usepackage{natbib}

\begin{document}

\title{Rectification by imprinted phase in a Josephson junction}

\author{G. R. Berdiyorov}
\affiliation{Departement Fysica, Universiteit Antwerpen,
Groenenborgerlaan 171, B-2020 Antwerpen, Belgium}
\author{M. V. Milo\v{s}evi\'{c}}
\affiliation{Departement Fysica, Universiteit Antwerpen,
Groenenborgerlaan 171, B-2020 Antwerpen, Belgium}
\author{L. Covaci}
\affiliation{Departement Fysica, Universiteit Antwerpen,
Groenenborgerlaan 171, B-2020 Antwerpen, Belgium}
\author{F. M. Peeters}
\email{francois.peeters@ua.ac.be}
\affiliation{Departement Fysica, Universiteit Antwerpen,
Groenenborgerlaan 171, B-2020 Antwerpen, Belgium}

\date{\today}

\begin{abstract}

A Josephson phase shift can be induced in a Josephson junction by a strategically nearby pinned Abrikosov vortex (AV). For an asymmetric distribution of imprinted phase along the junction (controlled by the position of the AV) such a simple system is capable of rectification of $ac$ current in a broad and tunable frequency range. The resulting rectified voltage is a consequence of the directed motion of a Josephson antivortex which forms a pair with the AV when at local equilibrium. The proposed realization of the ratchet potential by imprinted phase is more efficient than the asymmetric geometry of the junction itself, it is easily realizable experimentally, and provides rectification even in the absence of applied magnetic field.

\end{abstract}


\maketitle

Starting from the discovery of biological molecular motors \cite{astumian}, the ratchet effect has been demonstrated in many
different physical systems, where rectification in the presence of external random or periodic forces with zero time average is induced by means of spatial or temporal asymmetries (see Ref. \cite{review} for a review). Among other solid-state ratchet systems, superconducting ratchets have been realized - based on Abrikosov vortices \cite{aratchet}. Josephson vortex ratchets were also studied, in long Josephson junctions (JJs) \cite{Goldobin,Carapella2,Beck} and in specially engineered JJ arrays~\cite{Trias}. Voltage rectification based on the Josephson phase change has been demonstrated in asymmetric \cite{Weiss,Zapata} and three-junction SQUIDs~\cite{Sterck2,Sterck3}, and in annular JJs \cite{annular} with the asymmetric potential created by junction design, by inhomogeneous magnetic field \cite{Carapella2} or by extra current biasing \cite{Beck}. Josephson ratchets based on asymmetry of the drive rather than the potential itself have also been realized~\cite{Ustinov,Marchesoni}. The nonlinear signal mixing of two driving forces was also shown to be capable to control transport in different deterministic and Brownian ratchet devices \cite{nori,Machura2}.

In this Letter, we propose a ratchet based on an {\it inhomogeneous phase change} along the planar JJ. The simplest practical realization of such phase distribution can be realized by pinning an Abrikosov vortex (AV) nearby the junction (see Fig. 1). An AV can be inserted into the sample, e.g., by field cooling \cite{Hyun}, or by passing a large bias current through the system \cite{krasnov}. Once in the system, the location of the AV can be controlled by e.g. appropriately directed transport current \cite{Hyun}. However, such current would affect the JJ as well. A more elegant way of nucleating, as well as manipulating the AV is through the use of an electron beam, demonstrated experimentally by Ustinov {\it et al.} \cite{Ustinov2}. Here, we keep the AV from penetrating into the junction, since its magnetic field is known to strongly alter the properties of the junction \cite{Golubov}.

\begin{figure}[b]
\includegraphics[width=\linewidth]{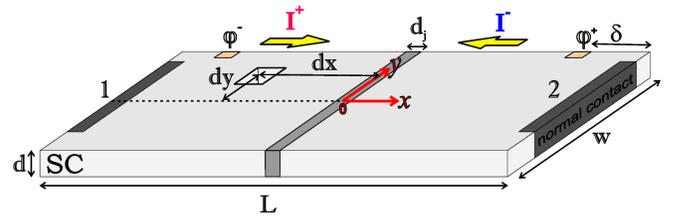}
\caption{\label{fig1}(color online) An oblique view of the system: a superconducting stripe (width $w$, length $L$ and thickness $d\ll\lambda, \xi$) with a central Josephson junction (width $d_j$). An Abrikosov vortex is trapped in a hole (of size $a$) located at a distance $dx$ from the junction ($y$-axis) and $dy$ from the $x$-axis. The current is applied via normal-metal contacts (labeled 1 and 2) and the output voltage is measured at a distance $\delta$ away from these leads.}
\end{figure}

Our idea is fairly simple. In the presence of an applied $ac$ current, an off-center location of the AV creates an asymmetric phase imprint on the junction and, consequently, an asymmetric potential for the motion of a Josephson fluxon along the junction. This in turn generates a net $dc$ voltage, i.e. rectification, in a broader current and frequency range than earlier ratchets based on geometric asymmetry of the junction. Moreover, a clear advance of the proposed ratchet is that it operates in the absence of applied magnetic field, whereas most known superconducting ratchets are not.

We consider a thin superconducting stripe with a narrow metallic junction (see Fig. \ref{fig1}). The location of the AV is predetermined by a hole
which acts as a pinning site. To take into account the Josephson coupling and the phase field of the AV inside the junction, we use the Ginzburg-Landau (GL) formalism with a Lawrence-Doniach extension \cite{Lawrence}. The resulting, modified time-dependent GL equation \cite{Kramer} takes the form
\begin{eqnarray}
&&\frac {u}{\sqrt{1+\Gamma^2|\psi|^2}}\left(\frac{\partial}{\partial
t}+i\varphi+\frac {\Gamma ^2}{2}\frac {\partial |\psi|^2}{\partial
t}\right)\psi=(\nabla-i\mathbf{A})^2\psi\nonumber\\
&&+\left(1-\frac{T}{T_c}-|\psi|^2\right)\psi+\frac{1}{\mu d_j^2}\left(\psi_{\pm d_j/2}- \psi_{\mp d_j/2}e^{\pm
i\mathbf{\bar{A}}}\right),\nonumber\\
\label{gleq}
\end{eqnarray}
coupled with the equation for the electrostatic potential $\Delta\varphi=\textrm{div}(\Im[\psi^*(\nabla-i\mathbf{A})\psi])$,
where {\bf A} is the magnetic vector potential. These coupled non-linear differential equations are solved self-consistently using Euler and multi-grid iterative procedures. The last term in Eq. (1) (with $\mathbf{\bar{A}}=\int_{-d_j/2}^{d_j/2}A_xdx$ and $\mu$ the ratio of the mass of the Cooper-pairs in the superconducting and normal-metal regions) describes the Josephson coupling across the junction of width $d_j$, and exists in the calculation just for the points bordering the junction on either side (indexed $\pm d_j/2$). Here, the order parameter is scaled to its value at zero magnetic field $\psi_0$, distances to coherence length $\xi(0)$, time to $\tau_{GL}=\pi\hbar/8k_BT_cu$, vector potential to $c\hbar/2e\xi(0)$, and the electrostatic potential to $\varphi_0=\hbar/2e\tau_{GL}$. The material parameter $\Gamma=2\tau_e\psi_0/\hbar$ (with $\tau_e$ being the inelastic scattering time) and $u$ are chosen to be 10 and 5.79, respectively \cite{Kramer}.
In the simulations we take $\xi(0)$=10 nm and $\lambda(0)=200$ nm, which are typical for Nb thin films \cite{Gubin}. Using the normal-state resistivity $\rho=18.7~\mu\Omega$cm for such films we obtain $\tau_{GL}\approx2.69$ ps and $\varphi_0=0.12$ mV at $T=0.9T_c$ which is the considered temperature in our simulations. Neumann boundary condition is applied at all sample boundaries (including the hole), except at the current contacts where we use $\psi=0$ and $\nabla\varphi|_n=-j$, with $j$ being the applied current density in units of $j_0=c\hbar/16e\pi^2\lambda^2\xi$.

\begin{figure} 
\includegraphics[width=\linewidth]{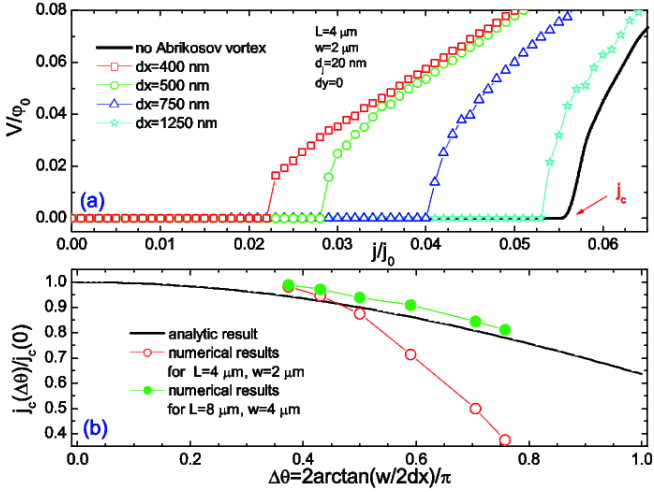}
\caption{\label{fig2}(color online) (a) $I$-$V$ characteristics of the sample with $L=4~\mu$m, $w=2~\mu$m and $d_j=20$ nm without (solid curve) and with the Abrikosov vortex (symbols) at a distance $dx$ away from the junction ($dy=0$). (b) The critical current $j_{c}$ vs. $\Delta \theta$ for two sizes of the sample (symbols) together with the analytical result (solid line, see text).}
\end{figure}

\begin{figure} 
\includegraphics[width=\linewidth]{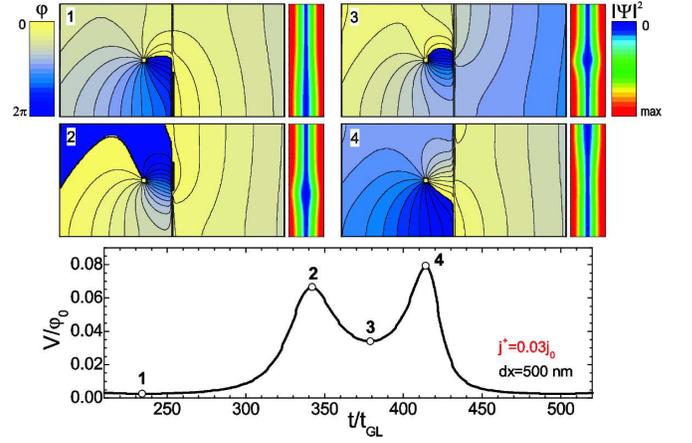}
\caption{\label{fig3}(color online) Voltage vs. time response of the sample at $j^{+}=0.03j_0$ for $dx=500$ nm, upon transition to the resistive state (see Fig. \ref{fig2}(a)). Panels (1-4) show snapshots of the phase (left) and density (right, just the junction area) of the order parameter at the times indicated in the $V(t)$ curve, and illustrate the motion of the Josephson vortex along the junction.}
\end{figure}

We begin our analysis by demonstrating the properties of a Josephson junction ($d_j=20$ nm) with symmetrically imprinted phase shift along the junction (i.e., $dy=0$). Figure \ref{fig2}(a) shows the time-averaged voltage vs. applied current ($I$-$V$) characteristics of the sample with length $L=4~\mu$m and width $w=2~\mu$m without (solid black curve) and with the AV (symbols) at different distance $dx$ from the junction. The $I$-$V$ curves exhibit a strong dependence on the vortex position -- the critical current $j_c$ for the transition to the resistive state decreases as the AV is placed closer to the junction and the imprinted phase difference $\Delta\theta=2\textrm{arctan}(w/2dx)/\pi$ increases. Fig. \ref{fig2}(b) shows the calculated $j_c$ as a function of $\Delta\theta$ (open dots) together with the known analytic expression for the critical current of the junction for given phase difference on the junction $j_{max}/j_c=\sin(\Delta\theta/2)/(\Delta\theta/2)$ (solid line). The latter showes good agreement with the recent experiments \cite{krasnov}. Although reasonably good agreement is found for smaller $\Delta\theta$ (i.e., larger distance $dx$), the numerical results strongly deviate from the analytical ones for close proximity of the AV to the junction. This faster decrease of $j_c$ with $\Delta\theta$ is due to the currents of the AV, which reach and interact with the junction interface. To eliminate this effect, we repeated the simulations for a sample with twice as large dimensions (i.e., for given $\Delta\theta$ the vortex is twice further from the junction), which indeed results in a $j_c$ very close to the analytic one (see filled dots in Fig. \ref{fig2}(b)). Small deviation from the analytical results is expected, due to the non-uniform phase distribution around the vortex and thus along the junction. Very recently Clem \cite{clem} studied analytically the effect of the nearby vortex on the critical parameters of a long planar Josephson junction, who reported several features of the system in common with a recent experiment \cite{krasnov}.

To show properties of the resistive state of the system, and the role of the imprinted phase therein, we plotted in Fig. \ref{fig3} the time evolution of the output voltage $V=\varphi^+-\varphi^-$ for an applied current just above $j_c$, for the sample with $dx=w/4$, together with snapshots of the distribution of phase and the modulus of the order parameter. The $V(t)$ characteristic shows periodic oscillations (Fig. \ref{fig3} shows one full period) with voltage peaks due to entry/exit of an ``antifluxon'' in the Josephson junction (see the insets of Fig. \ref{fig3}). This Josephson antivortex moves in the direction determined by the polarity of applied current, and during its motion forms a pair with the pinned AV, which leads to a local minimum in the $V(t)$ curve (see inset 3 in Fig. \ref{fig3}). Note that in the absence of the imprinted phase (i.e. pinned AV), the dissipation arises from the periodic nucleation and annihilation of the Josephson vortex-antivortex pairs (not shown here) in a similar fashion as in uniform superconducting stripes \cite{Weber}. In our case, the presence of the AV does not allow for the formation of a Josephson fluxon-antifluxon pair and only the antifluxon contributes to the resistive state of our sample, crossing the sample in opposite directions for changed polarity of the applied current.

\begin{figure} [t]
\includegraphics[width=\linewidth]{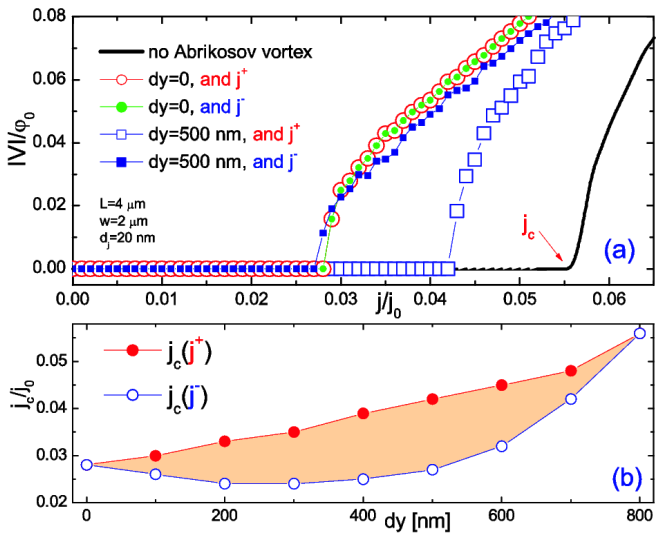}
\caption{\label{fig4}(color online) (a) $I$-$V$ curves of the sample of Fig. \ref{fig2} without (solid black curve) and with the pinned Abrikosov vortex (symbols) at $dx=500$ nm for different displacement $dy$ of the pinning site and for two opposite directions of the applied current. (b) The critical current $j_c$ for two directions of the external current as a function of $dy$ (for $dx=500$ nm).}
\end{figure}

In what follows, we show the dynamics of this antifluxon in an {\it asymmetrically} imprinted phase, obtained by shifting the AV off-center in the $y$-direction. Similarly to Fig. \ref{fig2}(a), Fig. \ref{fig4}(a) shows the $I$-$V$ curves, but now for two
different values of the shift $dy$ ($dx$ is fixed). In the symmetric case ($dy=0$, open and filled dots in Fig. \ref{fig4}(a)), $j_c$ for both directions of the applied current is the same (i.e., $j_c^+=j_c^-$) and the output voltage differs only by its sign (here we plot $|V|$) due to the reversal of the direction of the antifluxon motion. This situation changes entirely for $dy\neq 0$, i.e. for asymmetric distribution of imprinted phase - we observed a smaller needed (negatively polarized) current to induce the antifluxon motion along the junction (filled squares in Fig. \ref{fig4}(a)). The reason is the reduced energy barrier for the antifluxon entry, which is also reflected as a smaller entry peak in the periodic temporal characteristics of the voltage (shown in Supplementary material \cite{suppl1}). On the other hand, for the same reason $j_c$ increases for positive applied current (open squares in Fig. \ref{fig4}(a)), i.e., $j_c^+\neq j_c^-$. As a main result, we plotted in Fig. \ref{fig4}(b) the $j^+_c$ (filled circles) and $j^-_c$ (open circles) as a function of the asymmetry of imprinted phase (i.e., $dy$) for $dx=500$ nm. The difference in $j_c$ indicates that for a current amplitude in the shaded area of Fig. \ref{fig4}(b) only one direction for motion of the antifuxon is possible. In other words, when biased by an $ac$ current, our system
results in total {\it rectification} of the motion of the Josephson antivortex and the corresponding voltage. In our case, the ratchet potential is created by the imprinted phase field (thus not by geometry or temporal asymmetry in the junction, as is usually the case in vortex ratchets), where the phase distribution can be tuned by the position of the AV. For that reason, the rectified voltage, as well as the current range for rectification ($\Delta j=j_c^+-j_c^-$) strongly depend on the AV position (see Fig. \ref{fig4}(b)). The rectification ceases for a symmetric position of the AV ($dy=0$), which can be restored by a non-zero applied magnetic field.

\begin{figure} [t]
\includegraphics[width=\linewidth]{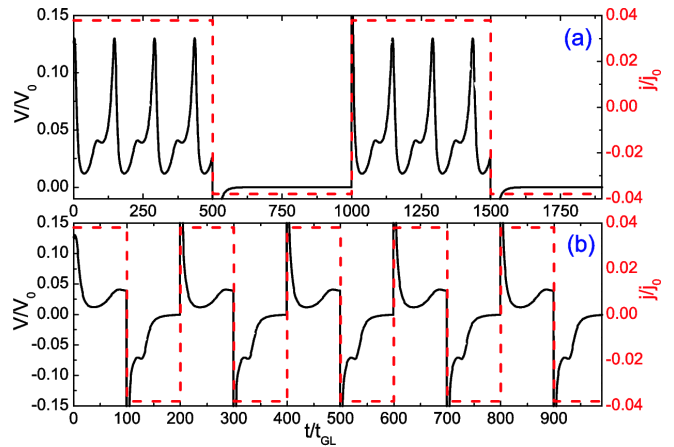}
\caption{\label{fig5}(color online) $V(t)$ characteristics (solid curves) of the sample with $dx=500$ nm and $dy=500$ nm, when biased by square-pulsed $ac$ current (dashed curves, right axes) with period $t_p=500~\tau_{GL}$ (a) and $t_p=100~\tau_{GL}$ (b).}
\end{figure}

To further demonstrate the operation of our ratchet system, we show its behavior in an $ac$ drive, i.e. a sequence of square current pulses of opposite polarity. Solid curve in Fig. \ref{fig5}(a) shows the calculated voltage vs. time for the amplitude $|j|=0.038j_0$ and period $t_p=500\tau_{GL}$ of the external current. Although the output voltage is rather weak (with an average $\sim10~\mu V$), the ratchet effect is clearly seen here. The averaged rectified voltage, which is roughly proportional to the antifluxon velocity, increases with increasing amplitude of the current, as also observed in the $I$-$V$ characteristics of the sample in Fig. \ref{fig4}(a). Negative voltage at the very moment of the current reversal is due to the inertial mass of the antifluxon which is already inside the junction. Since the voltage rectification in our system is a consequence of the directed net motion of the Josephson antifluxon, the rectification frequency (i.e., the frequency of the applied drive $f_r=1/t_p$ for which ratchet operates) is restricted by the characteristic time scale of the antifluxon dynamics -- its time needed to pass the entire junction, denoted as $\Delta t$. Therefore, if we drive our sample with square current pulses with period $t_p<\Delta t$, the ratchet behavior diminishes and the time averaged voltage becomes zero, as shown in Fig. \ref{fig5}(b) for $t_p=100~\tau_{GL}$ and $\Delta t=145~\tau_{GL}$. On a positive side, $\Delta t$ decreases with the amplitude of applied current, as well as with shortening the junction, so that the operating frequency range of our ratchet system {\it increases}. In the present calculation, for the chosen location of the AV with $dx=500$ nm and $dy=500$ nm, our ratchet rectifies square pulses with frequency up to 2 GHz, which is much larger than the frequency range of ratchets based on moving Abrikosov vortices \cite{aratchet}. We also tested our conclusions against sinusoidal current biasing, and obtained smaller rectification frequencies compared to square pulses with same amplitude, due to difficult temporal adjustment between the antifluxon motion and the driving force during the cycle. This fact is already well known experimentally (see e.g., Ref. \cite{Beck}).

\begin{figure} [t]
\includegraphics[width=\linewidth]{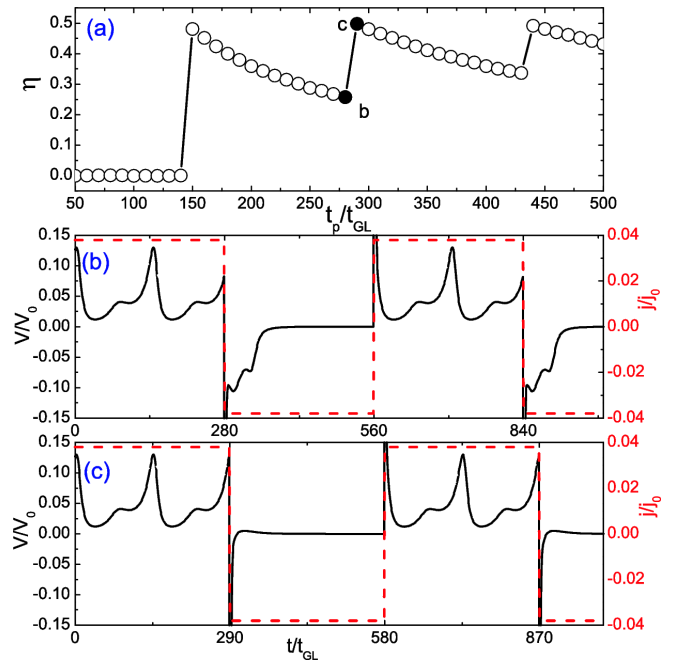}
\caption{\label{fig6}(color online) (a) The efficiency $\eta$ (defined in the text) of the phase-imprinted ratchet, as a function of the period $t_p$ of the applied square-pulsed current. (b,c) $V(t)$ characteristics for periods of the applied drive indicated in (a).}
\end{figure}

To quantify the efficiency of our ratchet system, and compare it to other ratchet realizations, we calculated the (long-)time-averaged voltage $\bar{V}^{ac}$ for the ac current drive relative to time-averaged voltage $\bar{V}^{dc}$ for a dc drive of same amplitude, as a function of the frequency of the square-pulsed ac current. The ratio $\eta=\bar{V}^{ac}/\bar{V}^{dc}$ is one of the common measures for the efficiency (see Ref. \cite{machura} and references therein for other definitions of ratchet efficiency), with a maximum of 0.5 for a temporally symmetric ac drive. As shown in Fig. \ref{fig6}(a), for $t_p>\Delta t$ (for $t_p<\Delta t$ rectification is not possible), our ratchet can have maximal theoretical efficiency which decreases to approximately 0.26 with decreasing frequency of the drive. Interestingly, with further decreasing the frequency, the maximal efficiency is {\it restored}, and the behavior of efficiency vs. frequency is {\it periodic}. This is a manifestation of commensurability between the period of the drive $t_p$ and the time for one Josephson crossing event $\Delta t$. As illustrated in Figs. \ref{fig6}(b,c), the maximal efficiency is obtained in our ratchet for $t_p=n\Delta t$ with $n$ an integer. For comparison, earlier Josephson ratchet proposals did not have such interesting commensurability effect, and also had lower efficiency, e.g. 0.33 in Ref. \cite{carapella3} or 0.22 in Ref. \cite{Ustinov}.

Finally, we discuss the efficiency of the phase imprint for the realization of a ratchet and compare it to the case of an asymmetric potential produced by varying the width of the JJ from narrow on one side, to wide on the other. The latter geometry should favor a higher voltage for the polarity of the current that drives the fluxon from the wide to the narrow end of the junction. However, in the absence of any applied magnetic field, we observe {\it identical voltage response} for both polarities of the applied current, i.e. no rectification is possible without external magnetic field (see Supplementary Material for animated data \cite{suppl2}). The reason is the generation of either fluxon or antifluxon in the resistive state with mirrored spatial dynamics for the two polarities of the applied current \cite{suppl2}. This directly shows that ratchet realization by imprinted phase is advantageous to earlier, geometry assisted vortex ratchets, in the absence of applied magnetic field.

In summary, we demonstrated a Josephson vortex rachet, based on a novel concept - imprinted asymmetric phase difference on a Josephson junction. The imprint of the phase can be realized by pinning the Abrikosov vortex in a suitable location nearby the junction, but even the direct phase imprint on the electronic condensate is not beyond experimental reach, as demonstrated on Bose-Einstein condensates \cite{phprint}. Our quantum ratchet rectifies voltage due to an applied $ac$ current with frequencies up to the GHz range (the frequency range is tunable by the response time of the sample), even at zero applied magnetic field. These properties, and its simple realization, make the phase-imprinted Josephson junction an advantageous fluxon diode in comparison to earlier proposed vortex ratchets.

This work was supported by the Flemish Science Foundation (FWO-Vlaanderen) and the Belgian Science Policy (IAP). G.R.B. and
L.C. acknowledge individual support from FWO-Vlaanderen.


\begin{thebibliography}{0}

\bibitem{astumian} F. J\"{u}licher, A. Ajdari, and J. Prost, Rev. Mod. Phys. {\bf 69}, 1269 (1997); R. D. Astumian, Science {\bf 276}, 917 (1997).

\bibitem{review} Special issue on \textit{Ratchets and Brownian Motors: Basics, Experiments and Applications} [Appl. Phys. A {\bf 75} (2002)]; P. H\"{a}nggi and F. Marchesoni, Rev. Mod. Phys. {\bf 81}, 387 (2009).

\bibitem{aratchet} F. Marchesoni, Phys. Rev. Lett. {\bf 77}, 2364 (1996); J. F. Wambaugh, C. Reichhardt, C. J. Olson, F. Marchesoni, and F. Nori, Phys. Rev. Lett. {\bf 83}, 5106 (1999); C. S. Lee, B. Jank\'{o}, I. Der\'{e}nyi and A. L. Barab\'{a}si, Nature (London) {\bf 400}, 337 (1999); C. J. Olson, C. Reichhardt, B. Jank\'{o}, and F. Nori, Phys. Rev. Lett. {\bf 87}, 177002 (2001); J. E. Villegas, S. Savel'ev, F. Nori, E. M. Gonzalez, J. V. Anguita, R. Garc\'{\i}a, and J. L. Vicent, Science {\bf 302}, 1188 (2003); B. Y. Zhu, F. Marchesoni, and F. Nori, Phys. Rev. Lett. {\bf 92}, 180602 (2004); J. Van de Vondel, C. C. de Souza Silva, B. Y. Zhu, M. Morelle, and V. V. Moshchalkov, Phys. Rev. Lett. {\bf 94}, 057003 (2005).

\bibitem{Goldobin} E. Goldobin, A. Sterck, and D. Koelle, Phys. Rev. E {\bf 63}, 031111 (2001); G. Carapella, Phys. Rev. B {\bf 63}, 054515 (2001); A. O. Sboychakov, S. Savel'ev, A. L. Rakhmanov, and F. Nori, Phys. Rev. Lett. {\bf 104}, 190602 (2010).

\bibitem{Carapella2} G. Carapella and G. Costabile, Phys. Rev. Lett. {\bf 87}, 077002 (2001).

\bibitem{Beck} M. Beck, E. Goldobin, M. Neuhaus, M. Siegel, R. Kleiner, and D. Koelle, Phys. Rev. Lett. {\bf 95}, 090603 (2005).

\bibitem{Trias} E. Tr\'{i}as, J. J. Mazo, F. Falo, and T. P. Orlando, Phys. Rev. E {\bf 61}, 2257 (2000); F. Falo, P. J. Mart\'{\i}nez, J. J. Mazo, and S. Cilla, Europhys. Lett. {\bf 45}, 700 (1999); K. H. Lee, Appl. Phys. Lett. {\bf 83}, 117 (2003); V. I. Marconi, Phys. Rev. Lett. {\bf 98}, 047006 (2007).

\bibitem{Zapata} I. Zapata, R. Bartussek, F. Sols, and P. H\"{a}nggi Phys. Rev. Lett. {\bf 77}, 2292 (1996).

\bibitem{Weiss} S. Weiss, D. Koelle, J. M\"{u}ller, R. Gross, and K. Barthel, Europhys. Lett. {\bf 51}, 499 (2000).

\bibitem{Sterck2} A. Sterck, R. Kleiner, and D. Koelle, Phys. Rev. Lett. {\bf 95}, 177006 (2005).
\bibitem{Sterck3} A. Sterck, D. Koelle, and R. Kleiner, Phys. Rev. Lett. {\bf 103}, 047001 (2009).

\bibitem{annular} A. Davidson, B. Dueholm, B. Kryger, and N. F. Pedersen, Phys. Rev. Lett. {\bf 55}, 2059 (1985); A.V. Ustinov, T. Doderer, R. P. Huebener, N. F. Pedersen, B. Mayer, and V. A. Oboznov, Phys. Rev. Lett. {\bf 69}, 1815 (1992); N. Gr{\o}nbech-Jensen, P. S. Lomdahl, and M. R. Samuelsen, Phys. Lett. A {\bf 154}, 14 (1991); A. Wallraff, Yu. Koval, M. Levitchev, M. V. Fistul and A. V. Ustinov, J. Low Temp. Phys. {\bf 118}, 543 (2000).

\bibitem{Ustinov} A. V. Ustinov, C. Coqui, A. Kemp, Y. Zolotaryuk, M. Salerno, Phys. Rev. Lett. {\bf 93}, 087001 (2004).

\bibitem{Marchesoni} F. Marchesoni, Phys. Lett. A {\bf 119}, 221 (1986); S. Flach, Y. Zolotaryuk, A. E. Miroshnichenko, and M. V. Fistul, Phys. Rev. Lett. {\bf 88}, 184101 (2002).

\bibitem{nori} S. Savel'ev, F. Marchesoni, P. H\"{a}nggi, and F. Nori, Phys. Rev. E {\bf 70}, 066109 (2004).

 \bibitem{Machura2} L. Machura, M. Kostur, and J. {\L}uczka, Chem. Phys. {\bf 375}, 445 (2010).

\bibitem{Hyun} O. B. Hyun, D. K. Finnemore, L. Schwartzkopf, and J. R. Clem, Phys. Rev. Lett. {\bf 58}, 599 (1987); O. B. Hyun, J. R. Clem, and D. K. Finnemore, Phys. Rev. B {\bf 40}, 175 (1989); Q. Li, J. R. Clem, and D. K. Finnemore, Phys. Rev. B {\bf 43}, 12843 (1991).

\bibitem{krasnov} T. Golod, A. Rydh, and V. M. Krasnov, Phys. Rev. Lett. {\bf 104}, 227003 (2010).

\bibitem{Ustinov2} A. V. Ustinov, T. Doderer, B. Mayer, R. P. Huebener, A. A. Golubov, V. A. Oboznov, Phys. Rev. B {\bf 47}, 944 (1993).

\bibitem{Golubov} A. A. Golubov and M. Y. Kupriyanov, Zh. Eksp. Teor. Fiz. {\bf 92}, 1512 (1987) [Sov. Phys. JETP {\bf 65}, 4 (1987)]; V. N. Gubankov, M. P. Lisitskii, I. L. Serpuchenko, F. N. Sklokin, and  M. V. Fistul, Supercond. Sci. Technol. {\bf 5}, 168 (1992); M. V. Fistul and G. F. Giuliani, Phys. Rev. B {\bf 58}, 9348 (1998); M. P. Lisitskiy and M. V. Fistul, Phys. Rev. B {\bf 81}, 184505 (2010).

\bibitem{Lawrence} W. E. Lawrence and S. Doniach, in \textit{Proceedings of the 12th International Conference on Low Temperature Physics}, Kyoto 1970, edited by E. Kanda (Keigaku, Tokyo, 1970), p. 361.

\bibitem{Kramer} L. Kramer and R. J. Watts-Tobin, Phys. Rev. Lett. {\bf 40}, 1041 (1978); R. J. Watts-Tobin, Y. Kr\"{a}henb\"{u}hl and L. Kramer, J. Low Temp. Phys. {\bf 42}, 459 (1981).

\bibitem{Gubin} A. I. Gubin, K. S. Il'in, S. A. Vitusevich, M. Siegel, and N. Klein, Phys. Rev. B {\bf 72}, 064503 (2005).

\bibitem{clem} J. R. Clem, arXiv:1109.1582v1 (unpublished).

\bibitem{Weber} A. Weber and L. Kramer, J. Low Temp. Phys. {\bf 84}, 289 (1991).

\bibitem{suppl1} Supplemental online video (EPAPS N0. xxx., http://www.aip.org/pubservs/epaps.html): Time evolution of the superconducting order parameter and the output voltage for the sample with asymmetric phase-imprint.

\bibitem{carapella3} Carapella, G. Costabile, N. Martucciello, M. Cirillo, R. Latempa, A. Polcari, G. Filatrella, Physica C {\bf 382}, 337 (2002).

\bibitem{machura} L. Machura, M. Kostur, P. Talkner, P. H\"{a}nggi, and J. {\L}uczka, Physica E {\bf 42}, 590 (2010).

\bibitem{suppl2} Supplemental online video (EPAPS N0. xxx., http://www.aip.org/pubservs/epaps.html): Time evolution of the superconducting order parameter and output voltage for the sample with asymmeteric Josephson junction and no phase imprint.

\bibitem{phprint} J. Denschlag, J.E. Simsarian, D.L. Feder, C.W. Clark, L.A. Collins, J. Cubizolles, L. Deng, E.W. Hagley, K. Helmerson, W.P. Reinhardt, S.L. Rolston, B.I. Schneider, and W.D. Philips, Science {\bf 287}, 97 (2000).

\end{thebibliography}
\end{document}